\documentclass[a4paper,12pt]{article}
\usepackage{epsfig}
\usepackage{epstopdf}\usepackage[sumlimits,intlimits,namelimits]{amsmath} 
\usepackage{dsfont}

\usepackage[english]{babel}

\usepackage{graphicx}

\usepackage{geometry}
\makeatletter
\def\fmslash{\@ifnextchar[{\fmsl@sh}{\fmsl@sh[0mu]}}
\def\fmsl@sh[#1]#2{%
  \mathchoice
    {\@fmsl@sh\displaystyle{#1}{#2}}%
    {\@fmsl@sh\textstyle{#1}{#2}}%
    {\@fmsl@sh\scriptstyle{#1}{#2}}%
    {\@fmsl@sh\scriptscriptstyle{#1}{#2}}}
\def\@fmsl@sh#1#2#3{\m@th\ooalign{$\hfil#1\mkern#2/\hfil$\crcr$#1#3$}}
\makeatother

\newcommand{\Li}[1]{\mathop{\mathrm{Li}}\nolimits_{#1}}
\newcommand{\ice}[1]{\relax}

\numberwithin{equation}{section}
\begin{document}
\begin{titlepage}
\begin{flushright}
SI-HEP-2016-15 \\[0.2cm]
QFET-2016-10
\end{flushright}

\vspace{1.2cm}
\begin{center}
{\Large\bf 
{\boldmath $B^0$-$\bar{B}^0$ \unboldmath} Mixing 
at Next-to-Leading Order}
\end{center}

\vspace{0.5cm}
\begin{center}
{\sc Andrey G. Grozin} \\[2mm] 
{\sf  Budker Institute of Nuclear Physics SB RAS, Novosibirsk 630090, Russia}  and \\
{\sf Novosibirsk State University, Novosibirsk 630090, Russia}
 \\[3mm] 
{\sc Rebecca Klein, Thomas Mannel } and {\sc Alexei A. Pivovarov} \\[0.1cm]
{\sf Theoretische Elementarteilchenphysik, Naturwiss.- techn. Fakult\"at, \\
Universit\"at Siegen, 57068 Siegen, Germany}
\end{center}

\vspace{0.8cm}
\begin{abstract}
\vspace{0.2cm}\noindent
We compute the perturbative corrections to the HQET 
sum rules for the matrix element of the $\Delta B=2$ operator 
that determines the mass difference of $B^0$, $\bar{B}^0$
states. Technically, we obtain analytically
the non-factorizable contributions 
at order $\alpha_s$ to the bag parameter that first appear at
the three-loop level.
Together with the 
known non-perturbative corrections due to vacuum condensates and 
$1/m_b$ corrections, the full next-to-leading order result 
is now available. We present a numerical value for the
renormalization group invariant bag parameter that 
is phenomenologically relevant 
and compare it with recent lattice determinations.    
\end{abstract}

PACS: 12.38.Bx, 12.38.Lg, 12.39.Hg, 14.40.Nd  
\end{titlepage}

\newpage
\pagenumbering{arabic}
\section{Introduction}
The mixing of states
in the systems of neutral flavored mesons belongs 
to the  most sensitive probes for effects from  
physics beyond the standard model (SM). While the mixing 
in the kaon and the charmed-meson 
systems has significant or even dominant long 
distance effect contribution, the mixing for the neutral 
$B$ mesons is dominated by the top-quark contribution and hence is 
dominated by short-distance physics. Technically, 
this fact means that the still necessary non-perturbative 
input is given by a matrix 
element of a local operator with $\Delta B = 2$, even if 
physics beyond the SM is present. 

Within the SM, the mixing frequency $\Delta m$ of 
the $B^0$--$\bar{B}^0$ oscillations
is determined by the following expression
\begin{equation} \label{DelM} 
\Delta m = \frac{G_F^2}{8 \pi^2} (V_{td}^* V_{tb})^2  F(x_t) m_t^2 
\eta_{\rm QCD}(\mu)   \langle B^0|Q(\mu)|\bar{B}^0 \rangle
\end{equation} 
where $x_t=m_t^2/m_W^2$, and 
\[
F(x) = \frac{1}{4} \left[ 1+ \frac{9}{1-x} - \frac{6}{(1-x)^2} 
- \frac{6 x^2}{(1-x)^3} \log x \right] 
\]   
is the Inami-Lim function~\cite{Inami:1980fz} (as a review, see, 
e.g.~\cite{Lenz:2006hd,Lenz,Nierste}).

The mass difference $\Delta m$ depends on 
the matrix element
$ \langle B^0|Q(\mu)|\bar{B}^0 \rangle$ of the local 
four-quark operator
\begin{equation}
Q = J_\mu J^\mu = Z(\alpha_s^{(n_f)}(\mu)) Q(\mu)\,,\quad
J^\mu = \bar{d}_L \gamma^\mu b_L\,,
\label{Intro:Q}
\end{equation}
where $b_L$, $d_L$ are the left-handed bare quark fields 
(see, e.\,g.,~\cite{BBL:96,Beneke:1998sy}). 
The 
short-distance coefficient $\eta_{\rm QCD}(\mu)$ 
in~(\ref{DelM}) accounts for contributions of scales larger than
the $b$-quark mass $m_b$.
The dependence of $\eta_{\rm QCD}(\mu)$
on the 
renormalization point $\mu$ compensates
the  $\mu$-dependence of the 
matrix element 
$ 
\langle B^0|Q(\mu)|\bar{B}^0
\rangle
$ 
that is the main object of low energy (for the scales down of $m_b$)
QCD analysis.
The matrix element of the four quark operator is 
traditionally written as
\begin{equation}
\langle B^0|Q(\mu)|\bar{B}^0 \rangle  =
2 \left(1 + \frac{1}{N_c}\right)
\langle B^0|J_\mu|0\rangle \cdot\langle0|J^\mu|\bar{B}^0\rangle
B(\mu)=2 \left(1 + \frac{1}{N_c}\right) f_B^2 M_B^2B(\mu)\, ,
\label{Intro:B}
\end{equation}
where $N_c$ is the number of colours, $N_c=3$ in QCD, $B(\mu)$ is the
bag parameter, and
\begin{equation}
\langle 0|J^\mu|\bar{B}^0(p) \rangle = - \frac{i}{2}   f_B p^\mu\,
\end{equation} 
is given by the $B$ meson decay constant $f_B$. 
Note that the decay constant $f_B$ is 
a physical quantity which is independent of the
renormalization point,
its numerical value is rather well known (as recent reviews, 
see, e.g.~\cite{Gelhausen:2013wia,Aoki:2016frl}).
Hence 
the full $\mu$ dependence
enters the bag parameter $B(\mu)$.

Setting $B(\mu) = 1$ corresponds to the naive factorization
prescription for the matrix element~(\ref{Intro:B})
which would be true for the bare operator 
$Q$ at tree level
but is spoiled by the strong interactions for the ``dressed''
operator $Q(\mu)$. 
The hadronic parameter $B(\mu)$ 
can only be obtained
by using some non-perturbative method,
such as lattice simulations (see, 
e.\,g.,~\cite{Aoki:2016frl,Aoki:2014nga,Gamiz:2009ku,Carrasco:2013zta,Lattice,Dowdall:2014qka})
or QCD sum 
rules~\cite{OP:88,Reinders:1988aa,KOPP:03,MPP:11,Pivovarov:2012zz}.
While the naive factorization estimate $B(m_B) = 1$ is rather
satisfactory even quantitatively, it is a kind of a model
assumption, and  
a key issue in the precision 
phenomenological analysis of the processes of
mixing is the determination of the deviation of $B(\mu)$
from unity. The matrix element appearing in (\ref{DelM}) 
still depends on $m_b$ which is a scale large 
compared to $\Lambda_{\rm QCD}$. To evaluate this matrix element 
further, we perform a 
heavy quark expansion (HQE) for this quantity, 
resulting in a combined  expansion in powers of 
$\alpha_s (m_b)$ and $\Lambda_{\rm QCD} / m_b$. 
The remaining matrix elements appearing in 
this expansion are defined in Heavy Quark Effective Theory (HQET) 
and may be estimated in an 
HQET sum rule. 

In a previous paper~\cite{MPP:11}, we have estimated the subleading 
terms of order 
$\Lambda_{\rm QCD} /m_b$ in such an expansion with 
a sum rule. However, in order to obtain the full next-to-leading order 
(NLO) result,  we also need to 
estimate the perturbative contributions of order $\alpha_s$. 
Within the framework of HQET sum rules 
this requires a calculation of three-loop diagrams.
The relevant master integrals have been found
in~\cite{GL:09}. In the present paper we give 
the results of the calculation for the bag 
parameter. With this calculation the complete NLO terms are now 
known. 

In the next section we collect some known perturbative results which
are 
needed to set up the sum rule 
calculation discussed in section~\ref{S:HQET}. Finally, 
we present a complete NLO result 
and discuss its implications for $B^0$--$\bar{B}^0$ mixing. 
 
\section{Perturbative Contributions to the Bag Parameter} 
In this section we collect some perturbation theory results 
relevant for the analysis of mixing.
 
The $\mu$ dependence of the bag parameter at scales above 
the $b$ quark mass is known to two 
loops~\cite{Buras:1990fn}, the result reads
\begin{align}
B(\mu) ={}& B(\mu_0)
\left(\frac{\alpha_s^{(n_f)}(\mu)}{\alpha_s^{(n_f)}(\mu_0)}\right)^{\gamma_0/(2\beta_0^{(n_f)})}
\Biggl[1 + \frac{\gamma_0}{2\beta_0^{(n_f)}} \left(
  \frac{\gamma_1}{\gamma_0} 
- \frac{\beta_1^{(n_f)}}{\beta_0^{(n_f)}} \right)
\frac{\alpha_s^{(n_f)}(\mu) - \alpha_s^{(n_f)}(\mu_0)}{4\pi}
\nonumber\\
&\hphantom{B(\mu_0)\left(\frac{\alpha_s^{(n_f)}(\mu)}
{\alpha_s^{(n_f)}(\mu_0)}\right)^{\gamma_0/(2\beta_0^{(n_f)})}\biggl[\Biggr.}
+ \mathcal{O}(\alpha_s^2) \biggr]
\nonumber\\
={}& \hat{B} \left(\alpha_s^{(n_f)}(\mu)\right)^{\gamma_0/(2\beta_0^{(n_f)})}
\left[1 + \frac{\gamma_0}{2\beta_0^{(n_f)}} \left(
    \frac{\gamma_1}{\gamma_0} 
- \frac{\beta_1^{(n_f)}}{\beta_0^{(n_f)}} \right)
\frac{\alpha_s^{(n_f)}(\mu)}{4\pi} + \mathcal{O}(\alpha_s^2) \right]\,,
\label{Intro:RG}
\end{align}
where the anomalous dimension of the operator $Q$ in (\ref{Intro:Q}) is
\begin{align}
&\gamma(\alpha_s) = \frac{d\log Z(\alpha_s(\mu))}{d\log\mu}
= \gamma_0 \frac{\alpha_s}{4\pi}
+ \gamma_1 \left(\frac{\alpha_s}{4\pi}\right)^2
+ \mathcal{O}(\alpha_s^3)\,,
\nonumber\\
&\gamma_0 = 6 \frac{N_c - 1}{N_c}\,,\quad
\gamma_1 = - \frac{N_c-1}{2 N_c}
\left( \frac{19}{3} N_c + 21 - \frac{57}{N_c} - \frac{4}{3} n_f \right)
\label{Intro:gamma}
\end{align}
where $n_f$ is the number of flavors including the $b$ quark. 
The $\beta$-function coefficients are 
\begin{equation}
\beta_0=\frac{11}{3}N_c-\frac{2}{3} n_f\, ,\quad
\beta_1 = \frac{34}{3} N_c^2 - \left(\frac{13}{3} N_c - \frac{1}{N_c}\right) n_f\, .
\label{Intro:beta}
\end{equation}
In the physical quantity $\Delta m$~(\ref{DelM}),
the $\mu$ dependence of $B(\mu)$ is compensated by the $\mu$ dependence
of the Wilson coefficient $F(x_t) \eta_{\text{QCD}}(\mu)$.

At scales $\mu$ below the $b$ quark mass the QCD operators 
are expanded into a series in   
$\Lambda_{\rm QCD} / m_b$ by employing HQET, see 
e.\,g.~\cite{N:94,MW:00,G:04}.  
In particular, the operator $Q$ in~(\ref{Intro:Q}) becomes~\cite{CFG:96,B:96}
\begin{equation}
Q(\mu) = 2 \sum_{i=1}^2 C_i(\mu) \tilde{Q}_i(\mu) +
\mathcal{O}\left(\frac{1}{m_b}\right)\, ,
\label{Intro:match}
\end{equation}
where the $1/m_b$ contributions have been discussed in~\cite{KM:92}. 
The leading order part is
\begin{align}
\tilde{Q}_1 &{}= \tilde{J}_{1\mu} \tilde{J}_2^\mu\,,\quad
\tilde{J}_1^\mu = \bar{d}_L \gamma^\mu h_+\,,\quad
\tilde{J}_2^\mu = \bar{d}_L \gamma^\mu h_-\,,
\label{Intro:Q1}\\
\tilde{Q}_2 &{}= \tilde{Q}_2' + \frac{1}{4} \tilde{Q}_1\,,\quad
\tilde{Q}_2' = \tilde{J}_1 \tilde{J}_2\,,\quad
\tilde{J}_1 = \bar{d}_L h_+\,,\quad
\tilde{J}_2 = \bar{d}_L h_- \, .
\label{Intro:Q2}
\end{align}
The bare field $h_+$ annihilates the HQET heavy quark (moving
with the four velocity $v$),
and $h_-$ creates the heavy antiquark (again moving with 
the four velocity $v$), 
which is a completely separate particle in HQET framework.
The factor two in~(\ref{Intro:match}) comes from the fact that 
there are two $b$ fields in $Q$,
one of them becomes $h_+$ and the other one $h_-$.
The HQET operators $\tilde{Q}_1,\tilde{Q}_2$
have opposite Fierz parities and hence don't 
mix under renormalization which is designed so to preserve 
 Fierz transformations.

The matrix elements of the leading HQET 
operators in~(\ref{Intro:Q1}), (\ref{Intro:Q2})
can be written as
\begin{align}
\langle {\bf B}^0|\tilde{Q}_1(\mu)|{\bf \bar{B}}^0 \rangle & =
\left(1 + \frac{1}{N_c}\right)\,
\langle {\bf B}^0|\tilde{J}_{2\mu}(\mu)|0 \rangle \,
\langle 0 |\tilde{J}_1^\mu(\mu)|{\bf \bar{B}}^0\rangle \,
\tilde{B}_1(\mu)\,,
\label{Intro:B1}\\
\langle {\bf B}^0|\tilde{Q}'_2(\mu)|{\bf \bar{B}}^0 \rangle &=
\left(1 - \frac{1}{2 N_c}\right)
\langle {\bf B}^0|\tilde{J}_2(\mu)|0 \rangle\,
\langle 0 |\tilde{J}_1(\mu)|{\bf \bar{B}}^0 \rangle 
\tilde{B}'_2(\mu)\,,
\label{Intro:B2}
\end{align}
where the $B$ meson states with a static $b$ quark $| {\bf B} \rangle$ 
are normalized non-relativistically
\begin{equation*}
\langle {\bf B} (p')| {\bf B} (p) \rangle = (2\pi)^3 \delta(\vec{p}\,'-\vec{p}\,)\,,\quad
|B(p) \rangle = \sqrt{2 p^0}\,| {\bf B}(p) \rangle + {\cal O} (1/m_b) \, ,
\end{equation*}
and
\begin{align*}
&\langle 0| \tilde{J}_1^\mu(\mu)|{\bf \bar{B}}^0 \rangle=
- \frac{1}{2} \langle 0|\tilde{\jmath}_1(\mu)|{\bf \bar{B}}^0 \rangle \,v^\mu\,,\quad
\langle 0 |\tilde{J}_1(\mu)|{\bf \bar{B}}^0 \rangle =
- \frac{1}{2}\langle 0 |\tilde{\jmath}_1(\mu)|{\bf \bar{B}} ^0\rangle \,,\\
&\langle{\bf B}^0| \tilde{J}_2^\mu(\mu) |0\rangle =
\frac{1}{2} \langle{\bf B}^0| \tilde{\jmath}_2(\mu) |0\rangle\,v^\mu\,,\quad
\langle{\bf B}^0| \tilde{J}_2(\mu) |0\rangle =
- \frac{1}{2} \langle{\bf B}^0| \tilde{\jmath}_2(\mu) |0\rangle\,,\\
&\tilde{\jmath}_1 = \bar{d} \gamma_5 h_+\,,\quad
\tilde{\jmath}_2 = \bar{d} \gamma_5 h_-\,,\\
&\langle 0| \tilde{\jmath}_1(\mu) |{\bf \bar{B}}^0\rangle = i F(\mu)\,,\quad
\langle{\bf B}^0| \tilde{\jmath}_2(\mu) |0\rangle = i F(\mu)\,.
\end{align*} 
The $B$ meson decay constant $\langle0|j^\mu|\bar{B}^0\rangle = i f_B p_B^\mu$
(where $j^\mu = \bar{d} \gamma_5 \gamma^\mu b$) is
\begin{equation}
f_B = \sqrt{\frac{2}{ m_B}}C(\mu) F(\mu) 
+ \mathcal{O}\left(\frac{1}{m_b}\right)\,,
\label{Intro:fB}
\end{equation}
where~\cite{EH:90}
\begin{equation}
j^\mu v_\mu = C(\mu) \tilde{\jmath}_1(\mu) 
+ \mathcal{O}\left(\frac{1}{m_b}\right)\,,\quad
C(m_b)=1 - 2 C_F \frac{\alpha_s(m_b)}{4\pi} + \mathcal{O}(\alpha_s^2)
\label{Intro:cFB}
\end{equation}
($C_F = (N_c^2-1)/(2 N_c)$).
The anomalous dimension of the operators $\tilde{\jmath}_{1,2}$ is~\cite{JM:91,BG:91,G:92}%
\footnote{The three-loop term is also known~\cite{CG:03}, but we don't need it.}
\begin{equation}
\tilde{\gamma}(\alpha_s) = - 3 C_F \frac{\alpha_s}{4\pi}
+ C_F \left[ \frac{2}{3} \pi^2 \left( C_A - 4 C_F \right)
+ \frac{1}{2} \left( 5 C_F - \frac{49}{3} C_A \right)
+ \frac{5}{3} n_l \right]
\left(\frac{\alpha_s}{4\pi}\right)^2 + \mathcal{O}(\alpha_s^3)\,,
\label{Intro:gammaj}
\end{equation}
where $n_l = n_f-1$ is the number of light flavors (now 
excluding $b$ quark),
and $C_A = N_c = 3$. In
terms of these parameters, 
the anomalous dimension of 
the operator $\tilde{Q}_1$ in~(\ref{Intro:Q1})~\cite{G:93}
can be written as
\begin{align}
&\tilde{\gamma}_1(\alpha_s) - 2 \tilde{\gamma}(\alpha_s)
= \delta_{11} \left(\frac{\alpha_s}{4\pi}\right)^2 
+ \mathcal{O}(\alpha_s^3)\,,
\nonumber\\
&
\delta_{11} = \frac{N_c-1}{3 N_c} \left[ 2 \pi^2 
\left( 3 N_c - 2 - \frac{6}{N_c} \right)
- 11  N_c^2 - 15 N_c - 12 + \frac{18}{N_c}
+ 2 ( N_c + 3) n_l \right]\,.
\label{Intro:gamma1}
\end{align}
Vanishing of the leading (linear in $\alpha_s$) term 
in~(\ref{Intro:gamma1})
reflects the (accidental) fact that at one loop and for scales below 
the $b$ quark mass, 
the naive factorization of the four quark operator $\tilde{Q}_1$
into a product of 
two bi-linear operators is scale independent, 
i.e.\ 
$\tilde{\gamma}_1 = 2 \tilde{\gamma}$~\cite{SV:88,PW:88}.
Therefore the $\mu$ dependence of $\tilde{B}_1(\mu)$ is weak and
contains no leading logarithms:
\begin{equation}
\tilde{B}_1(\mu) = \tilde{B}_1(\mu_0) \left[1 +
\frac{\delta_{11}}{2 \beta_0^{(n_l)}} 
\frac{\alpha_s^{(n_l)}(\mu) - \alpha_s^{(n_l)}(\mu_0)}{4\pi}
+ \mathcal{O}(\alpha_s^2) \right]\,.
\label{Intro:B1mu}
\end{equation}
The anomalous dimension of $\tilde{Q}_2$ is only known up 
to one loop order~\cite{CFG:96,B:96}:
\begin{equation}
\tilde{\gamma}_2(\alpha_s) - 2 \tilde{\gamma}(\alpha_s) 
= \delta_{20} \frac{\alpha_s}{4\pi}
+ \mathcal{O}(\alpha_s^2)\,,\quad
\delta_{20} = 4 \frac{N_c+1}{N_c}\,,
\label{Intro:gamma2}
\end{equation}
and therefore
\begin{align}
\tilde{B}_2(\mu)
&\equiv - \left(1 - \frac{1}{2 N_c}\right) \tilde{B}_2'(\mu) 
+ \frac{1}{4} \left(1 + \frac{1}{N_c}\right) \tilde{B}_1(\mu)\nonumber
\\
&= \tilde{B}_2(\mu_0)
\left(\frac{\alpha_s^{(n_l)}(\mu)}{\alpha_s^{(n_l)}(\mu_0)}\right)^{\delta_{20}/(2\beta_0^{(n_l)})}
\left[1 + \mathcal{O}(\alpha_s)\right]\,.
\end{align}
The matching to HQET is most conveniently performed at $\mu=m_b$,
such that the matching coefficients contain no large 
logarithms:
\begin{equation}
Q(m_b) = 2 \left( C_1(m_b) \tilde{Q}_1(m_b) + C_2(m_b) \tilde{Q}_2'(m_b) \right)
+ \mathcal{O}\left(\frac{1}{m_b}\right)\,,
\label{Intro:Match}
\end{equation}
where~\cite{CFG:96,B:96,FHH:91}
\begin{align}
&C_1(m_b) = 1 - \frac{8 N_c^2 + 9 N_c - 15}{2 N_c} \frac{\alpha_s^{(n_f)}(m_b)}{4\pi}
+ \mathcal{O}(\alpha_s^2)\,,
\nonumber\\
&C_2(m_b) = - 2 (N_c + 1) \frac{\alpha_s^{(n_f)}(m_b)}{4\pi}
+ \mathcal{O}(\alpha_s^2)\,.
\label{Intro:C12}
\end{align}
Taking the matrix element of~(\ref{Intro:Match}),
using~(\ref{Intro:B}), (\ref{Intro:B1}), (\ref{Intro:B2}),
and re-expressing $f_B$ via $F(m_b)$~(\ref{Intro:fB}),
we obtain
\begin{equation}
B(m_b) = \frac{C_1(m_b)}{C^2(m_b)} \tilde{B}_1(m_b)
- \frac{N_c - \frac{1}{2}}{N_c + 1} \frac{C_2(m_b)}{C^2(m_b)} \tilde{B}_2'(m_b)\,.
\label{Intro:Bmatch}
\end{equation}
Substituting $C_{1,2}(m_b)$~(\ref{Intro:C12}) and $C(m_b)$~(\ref{Intro:cFB}),
we arrive at
\begin{align}
B(m_b) ={}& \left[1 - \frac{4 N_c^2 + 9 N_c - 11}{2 N_c} 
\frac{\alpha_s^{(n_f)}(m_b)}{4\pi} \right] \tilde{B}_1(m_b)
+(2 N_c-1) 
\frac{\alpha_s^{(n_f)}(m_b)}{4\pi} \tilde{B}_2(m_b)
\nonumber\\
&{} + \mathcal{O}\left(\alpha_s^2,\frac{1}{m_b}\right)
\label{Intro:Bmb0}
\end{align}
where within the needed accuracy 
$\alpha_s^{(n_f)}(m_b) = \alpha_s^{(n_l)}(m_b)$.
Consequently, in order to obtain the QCD bag parameter
$B(\mu)$ with the NLO precision,
we only need the leading order $\tilde{B}_2$; 
in particular, we do not need the 
two-loop anomalous dimension of the operator $\tilde{Q}_2$. 

Dependence of $\tilde{B}_1(\mu)$ on $\mu$ is weak.
$\tilde{B}_1(m_b)$ is related to $\tilde{B}_1(\mu)$
(where $\mu$ is a low normalization point used in the sum rules)
by~(\ref{Intro:B1mu}).
Neglecting factorization breaking in the terms suppressed by $\alpha_s$,
i.\,e.\ setting $\tilde{B}_1(\mu)=\tilde{B}_2'(\mu)=1$ in these terms, we obtain
\begin{equation}
B(m_b) = \tilde{B}_1(m_b) - \frac{11}{2} \left(1 - \frac{1}{N_c}\right) \frac{\alpha_s(m_b)}{4\pi}\,.
\label{Intro:Bmb1}
\end{equation}
There are two sources of factorization violation in the QCD bag
parameter
$B(m_b)$:
the HQET bag
parameter $\tilde{B}_1$ of the matrix element of 
the HQET operator $\tilde{Q}_1$ (which will be considered in 
Sects.~\ref{S:HQET}, \ref{S:SR})
and the matching contribution~(\ref{Intro:Bmb1}).
As expected, they are suppressed as $1/N_c$ in the large $N_c$ limit.

This concludes the collection of necessary results concerning 
the renormalization of the matrix element of the 
four-quark operator and its matching to HQET at scales below the $b$
quark mass. 
The remaining task is to evaluate 
the hadronic matrix element  of the operator $\tilde{Q}_1$
in HQET, or the HQET bag
parameter $\tilde{B}_1$, for which we perform a sum-rule 
analysis in HQET using operator product expansion (OPE).

\section{OPE in HQET for sum rules}
\label{S:HQET}
In the following subsections we evaluate the matrix element of 
the four-quark operator $\tilde{Q}_1$
with HQET sum rules. 
We first consider the perturbative part of the sum rule, which 
requires a three-loop calculation of a suitably chosen 
correlator, and in a second step we study the quark-condensate 
contribution  to the HQET sum rule.    

\subsection{Leading Perturbative Part} 
To evaluate the matrix element, we use 
a vertex (three-point) correlation function that has been first
proposed for the analysis of the kaon 
mixing in ~\cite{Chetyrkin:1985vj}.
This correlator reveals the factorizable structure of the matrix
element more clearly than the two-point function
but is significantly more difficult to compute at NLO in QCD 
compared to the
calculation of the two-point function~\cite{Narison:1994zt}.
For the present analysis we however  
set up a three-point sum rule in
HQET where the computational difficulties have been 
solved~\cite{GL:09}. 
We consider the correlator
\begin{equation}
K = \int d^d x_1\,d^d x_2\,e^{i p_1 x_1 - i p_2 x_2} 
\langle 0 |T \tilde{\jmath}_2(x_2) 
\tilde{Q}_1(0) \tilde{\jmath}_1(x_1) | 0 \rangle 
\label{SR:K}
\end{equation}
of the operator $\tilde{Q}_1$ given in~(\ref{Intro:Q1}). 
Here we compute in dimensional regularization with 
$d = 4 - 2 \varepsilon$ dimensions. The currents
\begin{equation}
\tilde{\jmath}_1 = \bar{h}_+ \gamma_5 d\, ,\quad
\tilde{\jmath}_2 = \bar{h}_- \gamma_5 d\, .
\label{SR:j}
\end{equation}
interpolate the ground state of a static $B$ meson.  

Both the HQET quark and the HQET antiquark propagate only forward 
in time $x \cdot v$,
so that the product in~(\ref{SR:K}) is non-zero only 
at $x_1\cdot v<0$, $x_2\cdot v>0$
and thus the time-ordered product coincides with the product. 

The correlator $K$ depends on two scalar 
quantities $\omega_{1,2} = p_{1,2} \cdot v$,
$K=K(\omega_{1},\omega_{2})$
which correspond to the  
residual energies of the $b$ quark and the anti-$b$ quark respectively.  

\begin{figure}[ht]
\begin{center}
\hfill 
\includegraphics[scale=1.3]{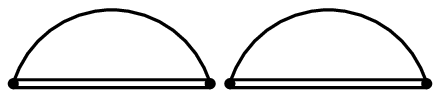}  \hfill 
\includegraphics[scale=1.3]{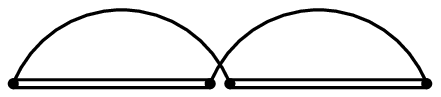} \hfill $\quad$
\end{center}
\caption{The leading perturbative contributions.
The currents $\tilde{J}_1$, $\tilde{J}_2$ are shown slightly split.}
\label{F:0}
\end{figure}

\begin{figure}[ht]
\begin{center} \hfill 
\includegraphics[scale=1.3]{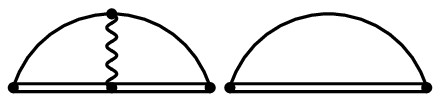}  \hfill
\includegraphics[scale=1.3]{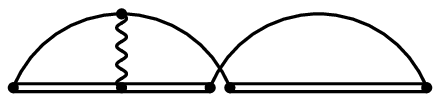} \hfill $\quad$ 
\end{center}
\caption{Some diagrams with corrections to the left loop.
Of course, similar corrections to the right loop exist.}
\label{F:1}
\end{figure}

The perturbative diagrams for the correlator $K$ can be 
subdivided into two classes.
The factorizable diagrams include the leading contributions (Fig.~\ref{F:0})
and those diagrams which contain corrections to the left loop
and to the right one separately (e.\,g., Fig.~(\ref{F:1})).
The right diagrams in Figs.~(\ref{F:0}, \ref{F:1}) are equal 
to the corresponding left diagrams
times the factor $(d-2)/(2 N_c)$.
This factor is obviously color suppressed $1/N_c$ at $d=4$:
there is one color loop ($N_c$) less,
and the Dirac structures can be reduced to products (as in the left diagrams)
by Fierz rearrangement.
At $d\ne4$ there is a contraction $\gamma_\mu\cdots\gamma^\mu$ within the same
$\gamma$-matrix string in each right diagram,
and it produces the factor $d-2$.

\begin{figure}[ht]
\begin{center} \hfill 
\includegraphics[scale=1.3]{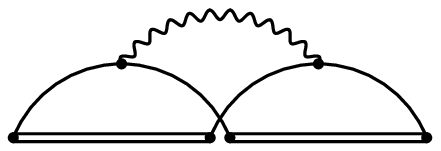} \hfill 
\includegraphics[scale=1.3]{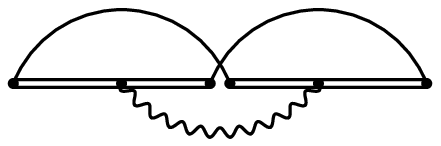} \hfill $\quad$ \\  \hfill 
\includegraphics[scale=1.3]{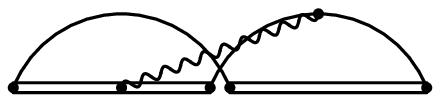} \hfill 
\includegraphics[scale=1.3]{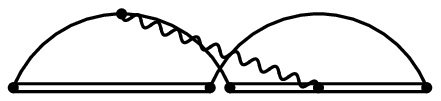} \hfill $\quad$ 
\end{center}
\caption{Nonfactorizable diagrams.}
\label{F:2}
\end{figure}

Nonfactorizable diagrams contain gluon exchanges between the left loop
and the right one.
They first appear at three loops (Fig.~\ref{F:2}).
Up to 3 loops, the results for the correlators $K(\omega_1,\omega_2)$ 
can be written as
\begin{equation}
K(\omega_1,\omega_2) = \left(1 + \frac{d-2}{2 N_c}\right)
\Pi(\omega_1) 
\Pi(\omega_2)
+ \Delta K(\omega_1,\omega_2)\,,
\label{Kstruct}
\end{equation}
where
\begin{equation}
\Pi(\omega) = \frac{N_c (-2\omega)^{2-2\varepsilon}}{(4\pi)^{d/2}} \left[ I_1
- 2 C_F \frac{g_0^2 (-2\omega)^{-2\varepsilon}}{(4\pi)^{d/2}} \frac{d-2}{d-4}
\left( I_1^2 - \frac{d (2d-5)}{d-4} I_2 \right) \right]
\label{Pi}
\end{equation}
is the correlator of $\tilde{\jmath}_1$ and 
$\tilde{J}_1$~\cite{BG:92,BBBD:92,N:92}, and
\begin{equation}
I_n = \Gamma(2n+1-nd) \Gamma^n\left({\textstyle\frac{d}{2}}-1\right)
\label{In}
\end{equation}
are the integrals corresponding to the ``sunset'' diagrams in HQET.  
The 3-loop nonfactorizable contribution is
\begin{equation}
\Delta K(\omega_1,\omega_2) = N_c C_F \frac{g_0^2}{(4\pi)^{3d/2}} 
R(\omega_1,\omega_2)\,.
\label{DK}
\end{equation}
We have reduced $R(\omega_1,\omega_2)$ to the master integrals 
investigated in~\cite{GL:09}
using the integration-by-parts program~\cite{L:12}
\begin{align}
R =&
- \frac{(d-2) (3d-7) (d^2-16d+40) (\omega_1 - 2 \omega_2)}{2 (d-4) (3d-8) \omega_1 (\omega_1 - \omega_2)}
I_3 (-2\omega_1)^{3d-5} + (\omega_1\leftrightarrow\omega_2)
\nonumber\\
&{} + \frac{(d-2) \bigl[(d-4) (3d-8) \omega_1 - (d-2) (2d-5) \omega_2\bigr]}
{(d-3) (d-4) \omega_1}
I_1 I_2 (-2\omega_1)^{2d-4} (-2\omega_2)^{d-3} + (\omega_1\leftrightarrow\omega_2)
\nonumber\\
&{} - \frac{(d-2) \bigl[(3d-8) (5d-14) \omega_1 - 2 (d-4) (d^2-7d+11) \omega_2\bigr]}
{(d-4) (3d-8) (\omega_1 - \omega_2)}
M_1(\omega_1,\omega_2) + (\omega_1\leftrightarrow\omega_2)
\nonumber\\
&{} + \frac{(d-2) (2d^2-15d+26)}{2 (d-3)} M_2(\omega_1,\omega_2)
+ \frac{(d-2)^2 \omega_1 \omega_2}{(d-3)^2} M_2'(\omega_1,\omega_2)
\nonumber\\
&{} + \frac{4 (d-2) (d-3) (d^2-16d+40) \omega_1 \omega_2}{(d-4) (3d-8)} M_3(\omega_1,\omega_2)
\nonumber\\
&{} - \frac{2 (d-2)^2 \omega_1}{d-4} M_4(\omega_1,\omega_2) + (\omega_1\leftrightarrow\omega_2)\,.
\label{R}
\end{align}

The next step is to expand the  master integrals 
around $d = 4$, i.e. in $\varepsilon$. The relevant technicalities are 
discussed in~\cite{GL:09} and in Appendix~\ref{S:A}. We obtain
\begin{equation}
\Delta K(\omega_1,\omega_2) =
N_c C_F \frac{g_0^2}{(4\pi)^{3d/2}} \left[\Gamma(1+2\varepsilon) 
\Gamma(1-\varepsilon)\right]^3
(-2\omega_1)^{2-3\varepsilon} (-2\omega_2)^{2-3\varepsilon} S(x)\,,
\label{Ke}
\end{equation}
where
\begin{equation}
x = \frac{\omega_2}{\omega_1}\,,
\label{x}
\end{equation}
and $S(x)=S(x^{-1})$ is
\begin{align}
S(x) =&
\left[\frac{1}{48} (x^2+x^{-2}) - \frac{\pi^2}{3} + \frac{5}{4}\right] 
\frac{1}{3 \varepsilon^2}
\nonumber\\
&{} + \left[- \frac{1}{16} (x^2-x^{-2}) \log x + \frac{61}{288}
  (x^2+x^{-2}) + x + x^{-1} - 4 \zeta_3 - \frac{4}{3} \pi^2 +
  \frac{41}{4} \right] 
\frac{1}{3 \varepsilon}
\nonumber\\
&{} + \frac{1}{2} \left(\frac{1}{16} (x^2+x^{-2}) + \frac{\pi^2}{3} - 
\frac{5}{4}\right) \log^2 x
- \left(\frac{61}{288} (x+x^{-1}) + 1\right) (x - x^{-1}) \log x
\nonumber\\
&{} + \frac{1}{216} \left(\pi^2 + \frac{2519}{24}\right) (x^2 + x^{-2})
- \frac{1}{3} \left(\frac{4}{9} \pi^2 - \frac{67}{4}\right) (x + x^{-1})
\nonumber\\
&{} - \frac{1}{3} \left(16 \zeta_3 + \frac{4}{45} \pi^4 + \frac{25}{6} \pi^2 - \frac{193}{4}\right)\,.
\label{S}
\end{align}

The correlator $K(\omega_1,\omega_2)$ is analytic at $\omega_{1,2}<0$.
It has a cut in $\omega_1$ from $0$ to $+\infty$ with the discontinuity
\begin{equation}
\rho_1(\omega_1,\omega_2) = \frac{1}{2\pi i}
\left[K(\omega_1+i0,\omega_2) 
- K(\omega_1-i0,\omega_2)\right]
\label{rho1}
\end{equation}
if we keep $\omega_2<0$.
The discontinuity  $\rho_1(\omega_1,\omega_2)$ as a function of
$\omega_2$ 
(at some $\omega_1>0$) has a 
cut from 0 to $+\infty$
with the discontinuity in $\omega_2$ 
\begin{equation}
\rho(\omega_1,\omega_2) = \frac{1}{2\pi i} 
\left[\rho_1(\omega_1,\omega_2+i0) - \rho_1(\omega_1,\omega_2-i0)\right]\,.
\label{rho}
\end{equation}
On dimensional grounds, the correlator at three loops has the form
\begin{equation}
K(\omega_1,\omega_2) = 
(-2\omega_1)^{2-3\varepsilon} (-2\omega_2)^{2-3\varepsilon} f(x)\,, 
\label{Kf}
\end{equation}
where the function $f$ can be gathered from the formulas given above. 
Looking at the spectral function 
$\rho_1 (\omega_1,\omega_2)$, we  
first  rotate $\omega_1$: we set $\omega_1=-\nu_1 e^{-i\alpha}$
($\nu_1>0$) 
and vary $\alpha$ from 0 to $\pi-0$ or $-\pi+0$
(keeping $\omega_2<0$); 
this gives
\begin{equation}
\rho_1(\nu_1,\omega_2) = \frac{(2\nu_1)^{2-3\varepsilon} 
(-2\omega_2)^{2-3\varepsilon}}{2\pi i}
\left[e^{3\pi i\varepsilon} f\left(-\frac{\omega_2}{\nu_1} e^{\pi i}\right)
- e^{-3\pi i\varepsilon} f\left(-\frac{\omega_2}{\nu_1} e^{-\pi i}\right) \right]\,,
\label{rho1f}
\end{equation}
where $\pi$ means $\pi-0$.
Now we set $\omega_2=-\nu_2 e^{-i\alpha}$ ($\nu_2>0$) and 
vary $\alpha$ from $0$ to $\pi-0$ or $-\pi+0$:
\begin{equation}
\rho(\nu_1,\nu_2) = \frac{(2\nu_1)^{2-3\varepsilon} (2\nu_2)^{2-3\varepsilon}}{(2\pi i)^2}
\left[ \left(e^{6\pi i\varepsilon} + e^{-6\pi i\varepsilon}\right)
  f(x) 
- f(x e^{2\pi i}) - f(x e^{-2\pi i}) \right]\,,\, 
x = \frac{\nu_2}{\nu_1}\,,
\label{rhof}
\end{equation}
where $x e^{\pm2\pi i}$ are at the Riemann sheets of 
the function $f(x)$ reached after crossing 
the cut in $x$ from $0$ to $-\infty$.

The bare double spectral density is
\begin{equation}
\rho(\omega_1,\omega_2) = \left(1 + \frac{1-\varepsilon}{N_c}\right) 
\rho(\omega_1) \rho(\omega_2)
+ \Delta \rho(\omega_1,\omega_2)\,,
\label{rho0}
\end{equation}
where~\cite{BG:92,BBBD:92,N:92}
\begin{equation}
\begin{split}
\rho(\omega) =& \frac{N_c (2\omega)^{2-2\varepsilon}}{(4\pi)^{d/2}}
\frac{\Gamma(1+2\varepsilon) \Gamma(1-\varepsilon)}{1-2\varepsilon}\\
&\left[1 + C_F \frac{g_0^2 (2\omega)^{-2\varepsilon}}{(4\pi)^{d/2}}
\Gamma(1+2\varepsilon) \Gamma(1-\varepsilon)
\left( \frac{3}{\varepsilon} + \frac{4}{3} \pi^2 + 17 \right) \right]\,,
\end{split}
\label{rhoBG}
\end{equation}
and
\begin{equation}
\Delta\rho(\omega_1,\omega_2) = N_c C_F \frac{g_0^2}{(4\pi)^{3d/2}}
\left[\Gamma(1+2\varepsilon) \Gamma(1-\varepsilon)\right]^3
(2\omega_1)^{2-3\varepsilon} (2\omega_2)^{2-3\varepsilon} r(x)\,,
\label{Dr}
\end{equation}
where $r(x)=r(x^{-1})$.
In the case of the operator $\tilde{Q}_1$ we have found that $r(x)$ 
does
not, in fact, depend on $x$
\begin{equation}
r(x) = - \left( \frac{4}{3} \pi^2 - 5 \right)\, .
\label{rx}
\end{equation}
The expression for $r(x)$ is a key computational result of our paper.

The renormalized double spectral density
$\rho_r(\omega_1,\omega_2) = 
\tilde{Z}_1^{-1} \tilde{Z}_j^{-2} \rho(\omega_1,\omega_2)$
is finite at the limit $\varepsilon\to 0$. This fact may be seen
explicitly by using (with 
$\alpha_s$ accuracy) the relation  $\tilde{Z}_1 = \tilde{Z}_j^2$ 
(see~(\ref{Intro:gamma1})).
Multiplying the factorizable part of~(\ref{rho0}) by 
$\tilde{Z}_1^{-1} \tilde{Z}_j^{-2} = \tilde{Z}_j^{-4}$
makes it finite separately.  Therefore, also the nonfactorizable part 
has to become finite separately.
At the limit $\varepsilon\to 0$ we obtain
\begin{equation}
\rho_r(\omega_1,\omega_2) = \left(1 + \frac{1}{N_c}\right) 
\rho_r(\omega_1) \rho_r(\omega_2)
+ \Delta\rho_r(\omega_1,\omega_2)\,,
\label{rhor}
\end{equation}
where~\cite{BG:92,BBBD:92,N:92}
\begin{equation}
\rho_r(\omega) = \frac{N_c (2\omega)^2}{(4\pi)^2} \left[ 1
+ C_F \frac{\alpha_s}{4\pi} \left( - 6 \log\frac{2\omega}{\mu} 
+ \frac{4}{3} \pi^2 + 17 \right) \right]
\label{rrBG}
\end{equation}
and
\begin{equation}
\Delta\rho_r(\omega_1,\omega_2) = - N_c C_F \frac{\alpha_s}{(4\pi)^5} 
(2\omega_1)^2 (2\omega_2)^2 
\left( \frac{4}{3} \pi^2 - 5 \right)\,.
\label{Drr}
\end{equation}
We note again, that for the operator $\tilde{Q}_1$ as given 
in (\ref{Intro:Q1}), $r(x)$ does not depend on $x$, i.e. on $\omega_{1,2}$; 
for other operators this is not necessarily so.

It is useful to rewrite the presentation~(\ref{Drr}) in the form
\begin{equation}
\Delta\rho_r(\omega_1,\omega_2) = -\frac{1}{ N_c} C_F \frac{\alpha_s}{4\pi} 
\rho_r(\omega_1) \rho_r(\omega_2)
\left( \frac{4}{3} \pi^2 - 5 \right)\,
\label{Drr-1}
\end{equation}
which is valid with $\mathcal{O}(a_s)$ accuracy. This form shows immediately the
deviation from the factorization with correct relative normalization
and can be used for the computation of corrections to the $B$ parameter.
Modifying the representation~(\ref{Drr-1})
even further one finds for the spectral density of three point
correlator at NLO
\[
\rho_r(\omega_1,\omega_2) = \left(1 + \frac{1}{N_c}\right) 
\rho_r(\omega_1) \rho_r(\omega_2)
+ \Delta\rho_r(\omega_1,\omega_2)
\]
\begin{equation}
= \left(1 + \frac{1}{N_c}\right) 
\rho_r(\omega_1) \rho_r(\omega_2)\left(1-\frac{\alpha_s}{4\pi} 
\frac{N_c-1}{2 N_c}\left( \frac{4}{3} \pi^2 - 5 \right)\right)\, 
\label{rhor-1}
\end{equation}
that is a master relation for the sum rules computation of ``direct''
contribution to 
$\Delta B$.

In the next subsection we compute 
the contributions of the quark condensate to 
the correlator~(\ref{SR:K}).  

\subsection{Quark Condensate Contribution}
\label{S:Q}
\begin{figure}[ht]
\begin{center} \hfill 
\includegraphics[scale=1.3]{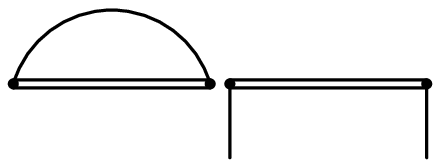}  \hfill 
\includegraphics[scale=1.3]{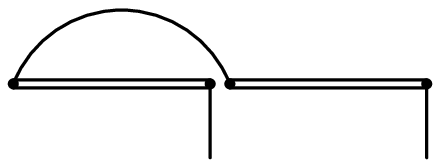}  \hfill $\quad$ 
\end{center}
\caption{The leading quark condensate contributions.
Of course, the mirror-symmetric diagrams also exist.}
\label{F:Q0}
\end{figure}

\begin{figure}[ht]
\begin{center} \hfill 
\includegraphics[scale=1.3]{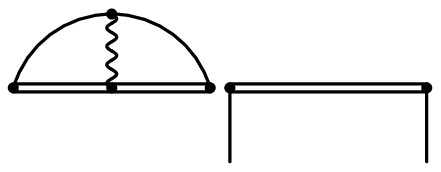} \hfill 
\includegraphics[scale=1.3]{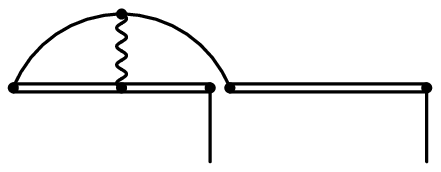}  \hfill $\quad$  \\  \hfill 
\includegraphics[scale=1.3]{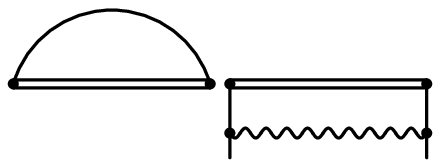} \hfill 
\includegraphics[scale=1.3]{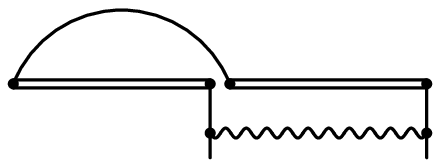}  \hfill $\quad$ 
\end{center}
\caption{Some of the factorizable contributions.}
\label{F:Q1}
\end{figure}

\begin{figure}[ht]
\begin{center}
\includegraphics{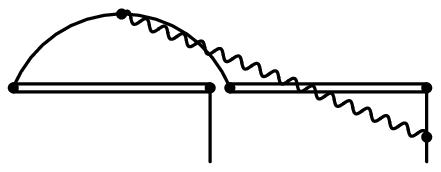} $\quad$ 
\includegraphics{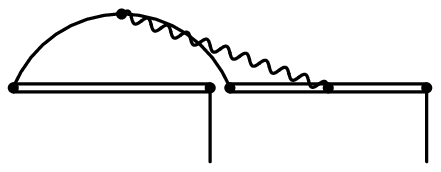} $\quad$ 
\includegraphics{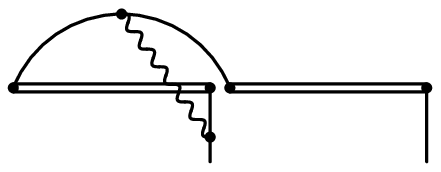} \\
\includegraphics{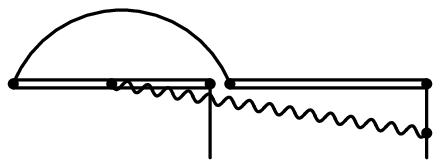} $\quad$ 
\includegraphics{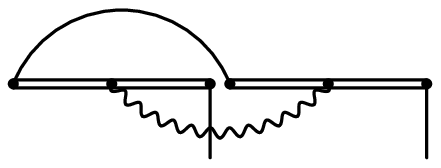} $\quad$ 
\includegraphics{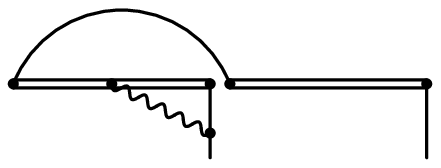} 
\end{center}
\caption{Nonfactorizable contributions
(the mirror-symmetric diagrams also exist).}
\label{F:Q2}
\end{figure}

The power correction to the sum rule discussed above are given 
in term of quark and gluon condensates. 
The leading term is given by the quark condensate contributions 
to the correlator $K$. The diagrams contributing to  
these power corrections are shown in Figs.~\ref{F:Q0}, \ref{F:Q1}, \ref{F:Q2}. 

The leading order quark condensate 
contribution (Fig.~\ref{F:Q0}) as well as some 
some of the 2-loop contributions (Fig.~\ref{F:Q1}) are factorizable.
They are contained in the product in~(\ref{Kstruct}), 
if we add the quark-condensate term~\cite{BG:92}
\begin{equation}
\Pi_q(\omega) = \frac{1}{2} \frac{\langle \bar{d}d\rangle}{-2\omega}
\left[1 + 2 C_F \frac{g_0^2 (-2\omega)^{-2\varepsilon}}{(4\pi)^{d/2}}
  (d-1) 
(d-4) I_1\right]\, .
\label{Piq}
\end{equation}
to the perturbative one~(\ref{Pi}).

The first nonfactorizable contributions due to quark condensate
appear at 
the two loop level as shown in Fig.~\ref{F:Q2}. 
The contribution of these diagrams to the correlator becomes
\begin{equation}
\Delta K_q(\omega_1,\omega_2) = C_F \frac{g_0^2
  {\langle }\bar{d}d{\rangle }}{(4\pi)^d} 
R_q(\omega_1,\omega_2)\,,
\label{DKq}
\end{equation}
where
\begin{align}
&R_q = \frac{4 (\omega_1+\omega_2) \left[(d-2) (d-5) 
(\omega_1^2+\omega_2^2) - (d^3 - 10 d^2 + 30 d - 30) \omega_1 \omega_2\right]}%
{(d-4) (-2\omega_1)^{5-d} (-2\omega_2)^{5-d}} I_1^2
\nonumber\\
&{} + \frac{2d-5}{2 (d-3) (d-4) (d-5) \omega_2^2 (\omega_1-\omega_2)}
\nonumber\\
&{} \times \bigl[ (d-2) (d-5)^2 \omega_1^3 + 2 (d-2) (d-5) (2d-5) 
\omega_1^2 \omega_2 - (d-3) (d^2 - 11 d + 6) \omega_1 \omega_2^2
\nonumber\\
&\qquad{} - 4 (d-2) (d-3) \omega_2^3 \bigr]
I_2 (-2\omega_1)^{2d-7} + (\omega_1\leftrightarrow\omega_2)
\nonumber\\
&{} + \frac{- (d-2) (d-5) \omega_1^3 - d \omega_1^2 \omega_2 
+ (d-3) (d-8) \omega_1 \omega_2^2 + (d-2) \omega_2^3}%
{4 (d-4) \omega_1 \omega_2^2 (\omega_1-\omega_2)}
M(\omega_1,\omega_2) + (\omega_1\leftrightarrow\omega_2)
\label{Rq}
\end{align}
where $M(\omega_1,\omega_2)$ is defined in~(\ref{App:M}).
Expanding in $\varepsilon$ we obtain
\begin{equation}
\Delta K_q(\omega_1,\omega_2) = C_F 
\frac{g_0^2 {\langle}\bar{d}d{\rangle}}{(4\pi)^d}
\left[ \Gamma(1+2\varepsilon) \Gamma(1-\varepsilon) \right]^2
(-2\omega_1)^{\frac{1}{2}-2\varepsilon} (-2\omega_2)^{\frac{1}{2}-2\varepsilon} S_q(x)\,,
\label{Kqe}
\end{equation}
where
\begin{align}
&S_q(x) = S_q(x^{-1}) = - \frac{7}{16} \frac{x^{1/2} + x^{-1/2}}{\varepsilon^2}
\nonumber\\
&{} + \biggl[ \frac{7}{2} (x^{1/2} - x^{-1/2}) \log x
+ (x^{1/2} + x^{-1/2}) (x + x^{-1} - 3) \frac{\pi^2}{3}
\nonumber\\
&\qquad{} - \frac{1}{4} (x^{1/2} + x^{-1/2}) (5 x + 5 x^{-1} + 14)
\biggr] 
\frac{1}{4\varepsilon}
\nonumber\\
&{} + (x^{1/2} + x^{-1/2}) (x + x^{-1} - 3) \left[ 3 \Li3(1-x) 
+ 3 \Li3(1-x^{-1}) - 2 L(x) \log x - 2 \zeta_3 \right]
\nonumber\\
&{} + (x^{1/2} - x^{-1/2}) (x + x^{-1}) L(x)
+ \frac{1}{8} (x^{1/2} + x^{-1/2}) (2 x + 2 x^{-1} - 7) \log^2 x
\nonumber\\
&{} + (x^{1/2} + x^{-1/2}) (10 x + 10 x^{-1} - 27) \frac{\pi^2}{24}
\nonumber\\
&{} + \frac{1}{8} (x^{1/2} - x^{-1/2}) (5 x + 5 x^{-1} + 32) \log x
- \frac{1}{4} (x^{1/2} + x^{-1/2}) (9 x + 9 x^{-1} + 11)\,.
\label{Sq}
\end{align}
Here the special function $L(x)$ is
\[
L(x)=-L(x^{-1})=\Li2(1-x)+\frac{1}{4} \log^2 x\, .
\]
Some useful properties of of this function and 
relevant polylogarithms ($\Li2$, $\Li3$) are given in the
Appendix.
 
Finally, the double discontinuity of the function 
$R_q(\omega_1,\omega_2)$ across the cuts $\omega_{1,2}>0$
reads
\begin{align}
&{\rm disc}_2~R_q(\omega_1,\omega_2) = 
2\Biggl[\left(\frac{\pi^2}{3}-\frac{5}{4}\right)
\omega_2^2\delta(\omega_1)
\nonumber\\
&-(\omega_2+\omega_1)\left(\frac{\omega_2}{\omega_1}
+\frac{\omega_1}{\omega_2}-3\right)
\log\left(1-\frac{\omega_1}{\omega_2}\right)
\Biggr]\theta(\omega_2-\omega_1)
\nonumber\\
&+(\omega_2\leftrightarrow\omega_1)\,.
\end{align}
Note that the coefficient of the 
$\delta(\omega_1)$ is related (up to a proportionality factor)
to that of a nonfactorizable perturbative correction
in eq.~(\ref{rx}).

The spectral density of quark condensate contribution now reads 
\begin{align}
&\Delta\rho_q(\omega_1,\omega_2) = C_F 
\frac{\alpha_s {\langle}\bar{d}d{\rangle}}{4\pi} 
\frac{2}{16\pi^2}\biggl\{
\nonumber\\
&\biggl[\left(\frac{\pi^2}{3}-\frac{5}{4}\right)
\omega_2^2\delta(\omega_1)
-(\omega_2+\omega_1)\left(\frac{\omega_2}{\omega_1}
+\frac{\omega_1}{\omega_2}-3\right)\log\left(1-\frac{\omega_1}{\omega_2}\right)
\biggr]\theta(\omega_2-\omega_1)
\nonumber\\
&+(\omega_2\leftrightarrow\omega_1)
\biggr\}\,.
\label{Drq-f}
\end{align}
The two-point correlator with the quark-condensate correction 
is given in~(\ref{Piq}).

\section{Sum Rules in HQET}
\label{S:SR}
The sum rule is now set up by comparing the perturbatively computed
correlator~(\ref{rhor-1})
with its hadronic representation. 
The hadronic spectral function is given by
\begin{equation}
\rho_H(\omega_1,\omega_2) = F^2\langle B |\tilde{Q}_1|B\rangle
\delta(\omega_1-\bar\Lambda)\delta(\omega_2-\bar\Lambda)
+\rho_{\rm cont}(\omega_1,\omega_2)
\label{rho-had}
\end{equation} 
where 
\begin{equation}
\langle \bar{B} |\tilde{Q}_1|B\rangle = 
\left(1+\frac{1}{N_c}\right)\frac{1}{4}F(\mu)^2 \tilde{B}_1 
= (1+1/N_c)\frac{1}{4}F(\mu)^2 (1+\Delta \tilde{B}_1 )
\label{rho-had-me}
\end{equation} 
and 
\begin{equation}
\rho_{\rm cont}(\omega_1,\omega_2)=\rho_{\rm PT}(\omega_1,\omega_2)
\left[1 - \theta(\omega_c-\omega_1) \theta(\omega_c-\omega_2)\right]\, .
\end{equation} 
Here $\bar{\Lambda}$ is the $B$ meson residual energy, 
$M_B - m_b = \bar{\Lambda}$  
and $\omega_c$ is the continuum threshold. 
One sees that if one considers also the sum rules for two point
correlators then the factorizable part of the matrix element
disappears and one has the direct prediction for $\Delta \tilde{B}_1$.

The simplest way to extract $\Delta \tilde{B}_1$ 
is to use the finite energy sum rules
(FESR) that equate the integrals over the
square $0<\omega_{1,2}<\omega_c$ of hadronic and OPE spectra.
One obtains for the perturbation theory 
contribution the folowing expression
\begin{equation}
\Delta\tilde{B}_1(\mu) =  - \frac{N_c-1}{2 N_c} 
\left( \frac{4}{3} \pi^2 
- 5 \right) \frac{\alpha_s^{(n_l)}(\mu)}{4\pi}
\approx  - 0.68 \frac{\alpha_s^{(n_l)}(\mu)}{\pi}
=-2.72\frac{\alpha_s^{(n_l)}(\mu)}{4\pi}\,.
\label{Bsr1}
\end{equation}
Here $n_l=4$. Eq.~(\ref{Bsr1}) gives a direct contribution to
the violation of factorization.

One can consider a more sophisticated analysis that controls power
corrections as in the Borel modification of dispersion sum rules. 
In HQET, however, there is a nice way
of solving the problem of controlling power corrections
suggested by 
the structure of dispersion representation for the correlators 
in configuration
space.
Indeed, in  
coordinate-space, the renormalized 
correlator~(\ref{SR:K}) at the parton level 
for Euclidean times $\tau_{1,2}$ ($\tau=i t$) becomes
\begin{equation}
K_r(\tau_1,\tau_2) = \int_0^\infty d\omega_1\,d\omega_2
\,e^{-\omega_1\tau_1-\omega_2\tau_2}\,\rho_r(\omega_1,\omega_2)
+  (\text{p.\,c.})\,,
\label{Ktau}
\end{equation}
where  (p.c.) represents the power corrections proportional 
to vacuum condensates. The power corrections are important mainly for
fixing the continuum threshold. We are not interested in the sum 
rules
analysis on its own but in precise determination of $\Delta {\tilde{B}}_1$.
Therefore we fix $\omega_c$ from all known sources (like $F(\mu)$ or $f_B$
eventually) and use the knowledge about two-point sum rules 
where the main
power correction is the quark condensate contribution.

The sum rule for the matrix element of the four-quark operator 
is obtained now from equating the OPE result to 
the hadronic expression for the correlator $K$ with the spectral 
density~(\ref{rho-had})
\begin{equation}
K_{\rm had}(\tau_1,\tau_2) = \int_0^\infty d\omega_1\,d\omega_2
\,e^{-\omega_1\tau_1-\omega_2\tau_2}\,\rho_{\rm had}(\omega_1,\omega_2)
\label{ktau-hadr}
\end{equation} 
which 
contains the desired matrix element~(\ref{rho-had-me}). With 
the usual duality assumption for the excited states, we obtain 
the sum rule 
\begin{equation}
F^2(\mu)\,\langle {\bf B} ^0|\tilde{Q}_1(\mu)|{\bf \bar{B}}^0 
\rangle \,e^{-\bar{\Lambda} (\tau_1+\tau_2)}
 = \int_0^{\omega_c} d\omega_1 \int_0^{\omega_c} d\omega_2\,
e^{-\omega_1\tau_1-\omega_2\tau_2}\,\rho_r(\omega_1,\omega_2)
+ (\text{p.\,c.})\,,
\label{sr3}
\end{equation}
with the same parameters 
$\bar{\Lambda}$, 
$M_B - m_b = \bar{\Lambda}$  
and the continuum threshold $\omega_c$. 
The Euclidean times $\tau_{1,2}$ ($\tau=i t$) play the role of
suppressing-higher-states parameters 
($1/\tau_{1,2}$ are the Borel parameters of the double Borel 
transform in $\omega_{1,2}$).
One can study the stability 
of the result with respect to varying $\tau_{1,2}$.
The version of sum rules 
in coordinate space in HQET is the most similar to the
lattice treatment of the problem.

Dividing the sum rule~(\ref{sr3})  
by two copies (product) 
of the two-point sum rules~\cite{BG:92,BBBD:92,N:92}
\begin{equation}
\frac{1}{2} F^2(\mu) e^{-\bar{\Lambda} \tau}
= \int_0^{\omega_c} d\omega\,e^{-\omega\tau}\,\rho_r(\omega) 
+ (\text{p.\,c.})\,,
\label{sr2}
\end{equation}
we finally obtain the result for the bag factor 
\begin{equation}
\tilde{B}_1(\mu) = 1 - \frac{N_c-1}{2 N_c} \left( \frac{4}{3} \pi^2 
- 5 \right) \frac{\alpha_s^{(n_l)}(\mu)}{4\pi} + (\text{p.\,c.})
\approx 1 - 0.68 \frac{\alpha_s^{(n_l)}(\mu)}{\pi} + (\text{p.\,c.})\,.
\label{Bsr}
\end{equation}
which coincides with that of the FESR approach.
This result is valid at a low normalization scale
$\mu\sim1/\tau_{1,2}$ or, in fact, $\mu\sim\omega_c$.
Also it assumes the same $\omega_c$ 
for both the two-point and three-point
correlators
(this is the reason why $\tilde{B}_1(\mu)$ 
is not explicitly dependent on $\omega_c$).
Thus, eq.~(\ref{Bsr}) gives the most complicated 
contribution to the bag parameter directly coming from
the three-loop correlation function (a ``direct'' violation of
factorization
to be contrasted with the violation in matching
given in Eq.~(\ref{Intro:Bmb1})).

There are still contributions originated from matching as given in 
Eq.~(\ref{Intro:Bmb1}) that should be added.
Let us add them first neglecting higher order corrections due to
different normalization points (running with NLO anomalous dimensions).
They give the total violation of factorization in the form 
\begin{equation}
-\frac{N_c - 1}{2 N_c} \left[ 11 \frac{\alpha_s(m_b)}{4\pi}
+ \left( \frac{4}{3} \pi^2 - 5 \right) \frac{\alpha_s(\mu)}{4\pi} \right]
\approx -(3.67+2.72) \frac{\alpha_s}{4\pi} .
\label{B-par0}
\end{equation}
where in the left-hand side we have still distinguished between 
the different scales
of $\alpha_s$ which appear on 
the one hand in the matching and on the other hand in 
the QCD sum rule. However, $\mu$ is not fixed 
and can be chosen somewhere in the vicinity of 
$\omega_c$ such that $\mu > \omega_c$.  
In our numerical analysis 
below we choose the scale to be $m_b$ and 
include the difference which is formally of order 
$\alpha_s(m_b)^2\log(m_b/\omega_c)$
in the uncertainty.  Nevertheless, 
one sees that the direct violation ($2.72$ in eq.~(\ref{B-par0}))
is quantitatively 
important and is comparable in magnitude with the violation in
matching ($3.67$ in eq.~(\ref{B-par0})).

The deviation of $\tilde{B}_1(\mu) $ from unity that we have found so
far measures the deviation 
from the naive factorization estimate due to perturbation theory
contribution to the OPE.  
Now we account for the contribution of quark condensate that violates
factorization. It can be important as its contribution to the
two-point sum rule that determines $F(\mu)$ and eventually $f_B$ is
not small.

After integrating the $\rho_q(\omega_1,\omega_2)$
within the finite energy sum rules one finds
\begin{equation}
\int\rho_q(\omega_1,\omega_2)d\omega_1d\omega_2
= C_F \frac{\alpha_s \langle\bar{q}q\rangle}{4\pi}
\frac{2}{3}\frac{\omega_c^3}{(4\pi)^2} \left(\pi^2 - \frac{149}{18}\right)\,.
\label{qq-sr}
\end{equation}
The two-point function sum rule~(\ref{sr2}) at $\tau=0$
(the finite-energy sum rule) gives
\[
m_B f_B^2 = 2 F^2 = N_c \frac{\omega_c^3}{3 \pi^2} - \langle\bar{q}q\rangle\,;
\]
we obtain
\begin{equation}
\Delta  \tilde{B}_1|_q = \frac{N_c - 1}{N_c}
\frac{\langle\bar{q}q\rangle}{m_B f_B^2}
\frac{\alpha_s}{4\pi}
\left[1 + \frac{\langle\bar{q}q\rangle}{m_Bf_B^2}\right]
\left(\pi^2-\frac{149}{18}\right)\,.
\label{qq-sr-fB}
\end{equation}
Numerically one has
\[
\frac{N_c-1}{N_c} \left(\pi^2 - \frac{149}{18}\right) \approx 1.06
\]
and 
\[
\frac{\langle \bar q q\rangle }{m_Bf_B^2}=-0.07
\]
for 
\[
\langle \bar q q \rangle =-(0.25~{\rm GeV})^3,
\quad m_B=5.3~{\rm GeV},
\quad  f_B = 200~{\rm MeV}
\]
that are typical values for the parameters.
In our numerical analysis we neglect the quark condensate contribution
in the square bracket in~(\ref{qq-sr-fB}).
One finds literally  
\begin{equation}
\Delta  \tilde{B}_1|_q=-0.08\frac{\alpha_s(m_b)}{4\pi}
\label{qq-sr-res}
\end{equation}
and after adding uncertainties we finally write
\begin{equation}
\Delta  \tilde{B}_1|_q=-(0.10\pm 0.04)\frac{\alpha_s(m_b)}{4\pi}\, .
\label{qq-sr-res-n}
\end{equation}
The contribution  
is rather small. Note that this is, in fact, a
numerical smallness.
Indeed, the result is 
a difference of two large numbers (of order 10)
$\left(\pi^2-\frac{149}{18}\right)\approx 9.9 - 8.3=1.6$ 
that happens to be small (of order 1).
Let us emphasize again that our estimates for the
phenomenological parameters have very generous uncertainties.
It is safe doing so because the contribution is rather small.

The nonPT terms (power corrections) have 
been analyzed in~\cite{OP:88} and then
extended and updated in~\cite{MPP:11}.
The FESR estimate from the latter is 
\begin{equation}  \label{NPTT}
\Delta B_{\rm cond}=-\frac{3\pi^2}{64}\left(\frac{1}{\omega_c^4}
\langle\frac{\alpha_s}{\pi} GG\rangle
-\frac{1}{\omega_c^5}\langle{\bar{q}Gq}\rangle\right)=
-\frac{3}{64}(0.06+0.1)=-0.008
\end{equation}  
for standard values of 
gluon condensate 
$\langle\frac{\alpha_s}{\pi} GG\rangle$~\cite{Novikov:1976tn} 
and mixed quark-gluon 
condensates $\langle{\bar{q}Gq}\rangle$
(e.g., see~\cite{Ovchinnikov:1988gk,Pivovarov:1991ie}).
The final result after an accurate Borel SR analysis in HQET 
reads for the $B_s$ meson~\cite{MPP:11}
\begin{equation}
\Delta B_{\rm cond}=-0.006\pm 0.005\, ,
\label{cond-lo-res}
\end{equation}
and we use this estimate also for the $B_d$ meson.

Because the values are very small they can be analyzed in linear
approximation that means that the consideration of sum rules with
included power corrections does not change the result for the parton
part (no mutual influence).

Nonfactorizable $1/m_b$ corrections can 
only emerge in the $\alpha_s/m_b$ order
(LO loops are completely factorized in QCD and this feature is
reproduced in HQET as well).
Therefore they are by factor 
$\Lambda/m_b=(0.5~{\rm GeV})/(5~{\rm GeV})=1/10$ smaller than
those analysed here and we simply include them in the uncertainty.
 
We discuss the final result in the next section where the comparison
with lattice is also given.

\section{Results and discussion} 
The main result of our analysis is the deviation $\Delta B$ from 
the value $B = 1$ in facorization.
In this section we collect all contributions and discuss the result. 

The partonic result (i.e. the purely perturbative contribution) 
consists of three pieces originating from the 
matching, from the QCD sum rule analysis and from the running: 
\begin{align*}
\Delta B|_{\rm PT} &= - \frac{N_c - 1}{2 N_c} \left[ 11 \frac{\alpha_s(m_b)}{4\pi}
+ \left( \frac{4}{3} \pi^2 - 5 \right) \frac{\alpha_s(\mu)}{4\pi} \right]
+ \frac{\delta_{11}}{2 \beta_0^{(n_l)}} \frac{\alpha_s(m_b) - \alpha_s(\mu)}{4 \pi}\\
&\approx - \left(\frac{4}{9} \pi^2 +2\right) \frac{\alpha_s}{4\pi}\, .
\end{align*}
As discussed after eq.~(\ref{B-par0}) we set   
for our numerical evaluation $\mu = m_b$ in the last step.   
Higher orders of $\alpha_s^2\log(m_b/\omega_c)$ 
can be taken through NLO anomalous 
dimension but they are small
and included as uncertainty in our analysis. 
To this end, we write
\[
\Delta B|_{\rm PT}=-6.4\frac{\alpha_s(m_b)}{4\pi}
\pm \left(X\frac{\alpha_s(m_b)}{4\pi}\right)\frac{\alpha_s(m_b)}{4\pi}
\]
where $X$ accounts for higher order terms. In order to estimate the uncertainty 
induced by such terms, we take a sizable value   
$X=20$ for this parameter, and we obtain
\[
\Delta B|_{\rm PT}=-6.4\frac{\alpha_s(m_b)}{4\pi}
\pm 0.3\frac{\alpha_s(m_b)}{4\pi}=
-(6.4\pm 0.3)\frac{\alpha_s(m_b)}{4\pi}\, .
\]
The choice of the value for the coupling constant 
is important for the
absolute estimate.
For the lattice estimates the reference value 
$
\alpha_s(M_Z)=0.1184
$
from~\cite{Bethke:2009jm} is usually used~\cite{Aoki:2016frl}.
Note that the estimate from the low energy $\tau$ decay data
gives a close value~\cite{Korner:2000xk}
\[
\alpha_s(M_Z)=0.1184\pm 0.0007|_{\text{exp}}\pm 0.0006|_{\text{hq mass}} \, .
\]
We stick, therefore, to the  
standard value
\begin{equation}
\alpha_s(m_{b})=0.20\pm 0.02
\label{alpha}
\end{equation}
with rather generous uncertainty
to account for possible systematic errors.

With the numerical value from~(\ref{alpha})
we obtain including systematic errors at the level of 30\%
\[
\Delta B_{\rm PT}=-0.10\pm 0.02\pm 0.03\, .
\]

We now turn to the non-perturbative condensate terms. The quark-condensate term 
computed in this paper at order $\alpha_s$ gives
\begin{equation}
\Delta B_q=-(0.10\pm 0.05)\frac{\alpha_s^{(n_l)}(m_b)}{4\pi}=-0.002\pm
0.001\, .
\label{qq-sr-res-1}
\end{equation}

In~\cite{MPP:11} the 
non-perturbative condensate terms that 
appear at tree level have been computed, 
see~(\ref{NPTT}). Their numerical value is~\cite{MPP:11}  
\[
\Delta B_{\rm nonPT}=-0.006\pm 0.005\, .
\]

Including everything we obtain the estimate 
\begin{equation}
\Delta B =-0.11 \pm 0.04\, 
\label{final-dB}
\end{equation}
where we summed errors in quadrature.

In order to compare this to other calculations, it is useful to employ 
the translation factor to the renormalization group invariant
parameter $\hat{B} = \hat{Z} B(m_b)$ is
\[
\hat{Z} =  \alpha_s(m_{b})^{-\frac{\gamma_0}{2\beta_0}}
\left(1 + \frac{ \alpha_s(m_{b})}{4\pi}
  \left(\frac{\beta_1\gamma_0-\beta_0\gamma_1}{2\beta_0^2}\right)\right)
\]
with
\[
\gamma_0 = 4, \quad \gamma_1 = -7 + \frac{4}{9}n_f,\quad n_f = 5,
\]
which numerically is
\[
\hat{Z}  = 1.51
\]
at $\alpha_s(m_b)=0.2$~\cite{Lattice}.

Applying this factor to our result 
\begin{equation}
B(m_b)|_{\rm this~paper} =1-(0.11\pm 0.04)
\label{final-Bmb}
\end{equation}
we obtain 
\begin{equation}
\hat B|_{\rm this~paper}= 1.51\left\{1-(0.11\pm 0.04)\right\}=1.34\pm 0.06\, .
\label{final-Bhat}
\end{equation}
The main uncertainty comes from the choice of scale for $\alpha_s(\mu)$
between $\mu\sim \omega_c$ and $m_b$, higher orders in
$\alpha_s(m_b)$, and the value of $\alpha_s(m_b)$.
The uncertainties due to other sources (like NNLO matching, or
systematics of sum rules) is difficult to quantify. For them we add
some typical values known from the experience with similar correlation
functions (see, e.g.~\cite{BBBD:92,N:92}).
More recent examples of uncertainty analysis within
sum rules approach can be found in~\cite{Gelhausen:2013wia,MPP:11}.   

We note that the sum rule yields a quite precise prediction. This is due to the 
fact, that the actual sum-rule calculation is performed for 
the deviation $\Delta B$ 
of the bag factor from unity. Although the calculation of   
$\Delta B$  suffers from 
the typical sum-rule uncertainty of tens of percents, the value 
obtained for $\hat{B}$ 
is quite precise since $\Delta B$ is small compared to unity. 
 
This value has to be compared to lattice value results.
The recent review~\cite{Aoki:2016frl} quotes the average
\[
\hat{B}_{\rm latt} = 1.26(9)
\]
for $n_f=2+1$ flavors based on~\cite{Aoki:2014nga,Gamiz:2009ku}
and
\[
\hat{B}_{\rm latt} = 1.30(6)
\]
for $n_f=2$~\cite{Carrasco:2013zta}.
The recent result~\cite{Lattice} is
\begin{equation}
\label{lattice-Bhat}
\hat{B}_{\rm latt} = 1.38(12)(6)
\end{equation}
The parameter $B$ itself normalized at the $b$ quark mass
is given earlier as~\cite{Dowdall:2014qka}
\[
{B}_{\rm latt}(m_b) = 0.8\pm 0.1
\]
(unfortunately, the number is not given explicitly and the result 
is extracted from the figure only).
At present, the progress in lattice computations 
is pretty fast and the results are going to further improve. Nevertheless, 
currently our sum rule estimate is competitive with the lattice calculations 
for the reasons discussed above. 

A comment on the QCD computation of the bag parameter with
the moments of the spectral density at the finite $b$-quark 
mass
used in 
the analysis of ref.~\cite{KOPP:03} is in order here.
The subtraction of divergences 
for the operator Q has been done in a way that is different
from the scheme adopted for the computation of the coefficient
functions of $\Delta B=2$ Hamiltonian in~\cite{Buras:1990fn}.
Thus, the renormalised operator $Q(\mu)$ 
of~\cite{KOPP:03} differs from the
one given in~\cite{Buras:1990fn} (and used in the present paper)
by a finite amount of
order $\alpha_s$. 
We are going to convert the results of~\cite{KOPP:03} to the
canonical basis in a separate paper.

\section{Summary} 
We have computed non-factorizable corrections to the bag parameter for
the $B_d^0-\bar{B}_d^0$ mixing.
The most complicated part is a ``direct'' contribution that requires
an acccount for three loop
diagrams in HQET. The main result of phenomenological analysis 
is that these corrections are small,
and factorization approximation is quantitatively valid.
We have found
\begin{equation}
B(m_b)-1=-(0.11\pm 0.04)
\label{B-final}
\end{equation}
and 
\begin{equation}
\hat B|_{\rm QCD}=1.34\pm 0.06\, 
\label{Bhat-final}
\end{equation}
for the $B_d$ meson bag parameter.

The main advantage of our approach is that we classify the
contributions (diagrams) at the level of OPE such that we can
explicitly single out contributions that completely factorize. In that
sense they can only produce unity in the bag parameter  
and do not require any computation if properly marked.
Subtracting these terms at the level of OPE we keep only terms that
explicitly violate factorization and use the sum rules for them.
It happens that those terms are numerically small and even
rather large uncertainties in their estimate still produce rather
precise result for the matrix element itself.

\subsection*{Acknowledgments}
We thank Th. Feldmann for the interest in the work and discussion.
A.G. is grateful to Siegen University for hospitality;
his work has been partially supported by the Russian Ministry of 
Education and Science.
This work is supported by the DFG Research Unit FOR 1873
"Quark Flavour Physics and Effective Theories".

\appendix
\section{Master integrals}
\label{S:A}

Expansions of the master integrals in $\varepsilon$ up to finite terms
have been obtained in~\cite{GL:09} Appendix~A.
However, we have found that the coefficients of $M_{3,4}$ in the correlator
are $\mathcal{O}(1/\varepsilon)$, and we need one more term in their expansions.
The expansion of $M_3$ is given by~(A.4) in~\cite{GL:09};
the new additional term in the braces is
\begin{align*}
&{} + \biggl[ 144 (2 x \log x - 1 + 19 x - 3 x^2) \Li3(1-x)
- 144 (2 x \log x + 3 - 19 x + x^2) \Li3(1-x^{-1})\\
&\qquad{} + 288 L^2(x)
+ 216 (1 - 7 x + x^2) L(x) \log x
+ 252 (1 - x^2) L(x)\\
&\qquad{} + \frac{81}{4} x \log^4 x
+ \frac{9}{2} (1 - x^2) \log^3 x
- \frac{9}{4} (19 + 70 x + 19 x^2) \log^2 x
+ 18 (1 - x^2) \log x\\
&\qquad{} - 8 \left( 630 \zeta_3 + \frac{71}{15} \pi^4 + 18 \pi^2 \right) x
+ 3 (11 - 120 x + 11 x^2)
\biggr] \varepsilon^4\,.
\end{align*}
The expansion of $M_4$ is given by~(A.5) in~\cite{GL:09};
the new additional term in the braces is
\begin{align*}
&{} - 2 \biggl[ 144 x^2 L_4(x)
- 12 x (2 x \log x + 3 + 18 x - 3 x^2) \Li3(1-x)\\
&\qquad{} + 12 x (2 x \log x - 1 - 18 x + x^2) \Li3(1-x^{-1})\\
&\qquad{} - 24 x^2 L^2(x)
+ 6 x \left[ 4 x \log^2 x + 18 x \log x - 5 (1 - x^2) \right] L(x)\\
&\qquad{} + 3 x (1 - x^2) \log^3 x
+ x \left[ 8 \pi^2 x + 3 (5 - 9 x - 5 x^2) \right] \log^2 x\\
&\qquad{} + 3 x \left[ 4 (8 \zeta_3 + 3 \pi^2) x - 1 + x^2 \right] \log x\\
&\qquad{} + 2 \left( 270 \zeta_3 + \frac{28}{15} \pi^4 + 9 \pi^2 \right) x^2
+ 2 x (7 + 2 x - x^2)
\biggr] \varepsilon^4\,,
\end{align*}
where the function
\[
L_4(x) = - L_4(x^{-1}) = \Li4(x)
+ \frac{1}{6} \log^3 x \log(1-x)
- \frac{1}{16} \log^4 x
- \frac{\pi^2}{12} \log^2 x
- \frac{\pi^4}{90}
\]
is analytical in $(0,+\infty)$
(no branching singularity at $x=1$).
We have also checked that the expansions~(A.2) and~(A.3) of $M_2$, $M_2'$
in~\cite{GL:09} satisfy the identity
\[
M_2' = \frac{d-3}{(d-4) \omega_1^2 \omega_2}
\left[ (\omega_1^2 - \omega_2^2) \frac{\partial M_2}{\partial \omega_2}
+ \frac{1}{2} (3d-8) (\omega_1 + 2 \omega_2) M_2 \right]
\]
following from IBP.

\begin{figure}[h]
\begin{center}
\includegraphics{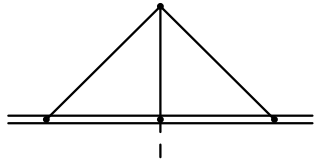}
\end{center}
\caption{Topology of 2-loop integrals}
\label{F:L2}
\end{figure}

For the calculation of 2-loop diagrams in Sect.~\ref{S:Q}
we need Feynman integrals shown in Fig.~\ref{F:L2}.
Using LiteRed~\cite{L:12} we reduce them to 3 trivial master integrals
$I_1^2 (-2\omega_1)^{d-3} (-2\omega_2)^{d-3}$, $I_2 (-2\omega_1)^{2d-5}$, $I_2 (-2\omega_2)^{2d-5}$
and 2 nontrivial ones,
\begin{equation}
M(\omega_1,\omega_2) = \raisebox{-3.5mm}{\includegraphics{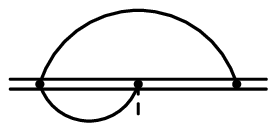}} = I_1 I(3-d,1,1;\omega_1,\omega_2)
\label{App:M}
\end{equation}
and $M(\omega_2,\omega_1)$.
Expansion of $M(\omega_1,\omega_2)$ in $\varepsilon$ is
\begin{align*}
&M(\omega_1,\omega_2) = -
\frac{\Gamma^2(1-\varepsilon) \Gamma(1+4\varepsilon)}{16 \varepsilon^2 (1-2\varepsilon) (1-4\varepsilon) (3-4\varepsilon)}
\Bigl\{ x (x-1) - (4 x^2 - 6 x + 1) \varepsilon\\
&{} - 2 \bigl[ x (x-1) \bigl( 4 L(x) + \log^2 x \bigr) - 2 (2x-1) \log x \bigr] \varepsilon^2\\
&{} + 8 \bigl[ x (x-1) \bigl( 4 \Li3(1-x) + 2 \Li3(1-x^{-1}) - 4 L(x) \log x - \tfrac{1}{3} \log^3 x + 4 L(x) \bigr)\\
&\qquad{} + (x^2 + x - 1) \log^2 x \bigr] \varepsilon^3 + \cdots \Bigr\} \frac{(-2\omega_2)^{2-4\varepsilon}}{x^2}\,.
\end{align*}

For calculations of spectral densities we used
\begin{align*}
&\Li2(1-x e^{\pm2\pi i}) = 
\Li2(1-x) \mp 2 \pi i \left[ \log|x-1| \pm \pi i \theta(x-1) \right]\,,\\
&\Li3(1-x e^{\pm2\pi i}) = 
\Li3(1-x) \mp \pi i \left[ \log|x-1| \pm \pi i \theta(x-1) \right]^2\,,\\
&\Li{n}(x+i0) - \Li{n}(x-i0) = \frac{2\pi i}{\Gamma(n)} \log^{n-1} x\quad (x>0)
\end{align*}
(where $1-x e^{\pm2\pi i}$ are on the Riemann sheets reached after crossing the cut).
We also used the identity
\[
\Li3(x) + \Li3(1-x) + \Li3(1-x^{-1}) = \frac{1}{6} \log^3 x - \frac{1}{2} \log^2 x\,\log(1-x) + \frac{\pi^2}{6} \log x + \zeta_3\,.
\]

\end{document}